\documentclass[usenatbib,usegraphicx]{mn2e}

\usepackage{graphicx}
\usepackage{txfonts}
\usepackage{natbib}
\usepackage[usenames,dvipsnames]{color}
\usepackage{pdflscape}
\usepackage{multirow}
\usepackage[colorlinks=true,citecolor=blue]{hyperref}
\newcommand{\nodata}{\ensuremath{\cdots}}
\newcommand{\igrheavy}{IGR~J17544$-$2619}
\newcommand{\software}[1]{\texttt{\small \MakeUppercase{#1}}}
\def \inte {INTEGRAL}

\def \sax {{\em BeppoSAX}}
\newcommand{\nustar}{{\em NuSTAR}}
\def \suzaku {{\em Suzaku}}
\def \sw {{\em Swift}}
\def \xrt {{\em XRT}}

\def \ferg {erg cm$^{-2}$ s$^{-1}$}
\def \ATel {ATel}

\newcommand\apj{ApJ}	
\newcommand\apss{Ap\&SS}	
\newcommand\aap{A\&A}	
\newcommand\baas{BAAS}	
\newcommand\mnras{MNRAS}	

\title[Magnetic field in \igrheavy]{\nustar\ Detection Of A Cyclotron Line In The Supergiant Fast X-ray Transient IGR~J17544$-$2619}

\author[Bhalerao et al.]{Varun Bhalerao$^{1}$\thanks{Email: varunb@iucaa.ernet.in (VB)}, 
Patrizia Romano$^{2}$, 
John Tomsick$^{3}$, 
Lorenzo Natalucci$^{4}$, 
David M. Smith$^{5}$,  \newauthor
Eric Bellm$^{6}$, 
Steven E. Boggs$^{3}$, 
Deepto Chakrabarty$^{7}$, 
Finn E. Christensen$^{8}$,  \newauthor
William W. Craig$^{3,9}$, 
Felix Fuerst$^{6}$, 
Charles J. Hailey$^{10}$, 
Fiona A. Harrison$^{6}$,  \newauthor
Roman A. Krivonos$^{3}$, 
Ting-Ni Lu$^{6,11}$, 
Kristin Madsen$^{6}$, 
Daniel Stern$^{12}$, 
George Younes$^{13}$, \newauthor
and 
William Zhang$^{14}$ \\
$^{1}${Inter University Centre for Astronomy and Astrophysics, P. O. Bag 4, Ganeshkhind, Pune 411007, India} \\
$^{2}${INAF, Istituto di Astrofisica Spaziale e Fisica Cosmica, Via U.\ La Malfa 153, I-90146 Palermo, Italy} \\
$^{3}${Space Sciences Laboratory, University of California, Berkeley, CA 94720, USA} \\
$^{4}${Istituto Nazionale di Astrofisica, INAF-IAPS, via del Fosso del Cavaliere, 00133 Roma, Italy; } \\
$^{5}${Physics Department and Santa Cruz Institute for Particle Physics, University of California, Santa Cruz, 1156 High St., Santa Cruz, CA 95064 USA} \\
$^{6}${Cahill Center for Astronomy and Astrophysics, Caltech, Pasadena, CA 91125} \\
$^{7}${Kavli Institute for Astrophysics and Space Research, Massachusetts Institute of Technology, 70 Vassar Street, Cambridge, MA 02139, USA} \\
$^{8}${DTU Space, National Space Institute, Technical University of Denmark, Elektrovej 327, DK-2800 Lyngby, Denmark} \\
$^{9}${Lawrence Livermore National Laboratory, Livermore, CA 94550, USA} \\
$^{10}${Columbia Astrophysics Laboratory, Columbia University, New York, NY 10027, USA} \\
$^{11}${Institute of Astronomy, National Tsing Hua University, Taiwan} \\
$^{12}${Jet Propulsion Laboratory, California Institute of Technology, Pasadena, CA 91109, USA} \\
$^{13}${USRA, NSSTC, 320 Sparkman Drive, 35801, Huntsville, Al., USA} \\
$^{14}${NASA Goddard Space Flight Center, Greenbelt, MD 20771, USA}
}

\begin{document}

\date{Accepted ---. Received ---; in original form ---}

\pagerange{\pageref{firstpage}--\pageref{lastpage}} \pubyear{X}

\maketitle

\label{firstpage}

\begin{abstract}
We present \nustar\ spectral and timing studies of the Supergiant Fast X-ray Transient (SFXT) \igrheavy. The spectrum is well-described by a $\sim1$~keV blackbody and a hard continuum component, as expected from an accreting X-ray pulsar. We detect a cyclotron line at 17~keV, confirming that the compact object in \igrheavy\ is indeed a neutron star. This is the first measurement of the magnetic field in a SFXT. The inferred magnetic field strength, $B = (1.45\pm0.03) \times 10^{12}~G (1 + z)$ is typical of neutron stars in X-ray binaries, and rules out a magnetar nature for the compact object. We do not find any significant pulsations in the source on time scales of 1--2000~s.
\end{abstract}


\section{Introduction}\label{sec:intro}

High mass X--ray binaries (HMXBs) are stellar systems composed 
of a compact object (either a neutron star or a black hole) 
and an early-type non-degenerate massive star primary.  
These systems are traditionally divided in two sub-classes 
\citep[e.g.][and references therein]{Reig2011}, 
depending on the nature of the primary that acts as as a mass donor, 
and the mass-transfer and accretion mechanisms onto the  compact object. 
While the Be/X--ray binaries (BeXBs) have main sequence Be star primaries, 
and are only observed as transient sources showing bright outbursts 
lasting a few days, the OB supergiant binaries (SGXBs) are persistent 
systems with an evolved OB supergiant primary.

%
Among the $\sim 250$ HMXBs known to populate our Galaxy and the Magellanic Clouds 
\citep[][]{Liu2005:hmxb_LMC_SMC,Liu2006:hmxb_Gal} 
a relatively small class termed Supergiant Fast X-ray Transients 
was recently recognized that  
shares properties with both BeXBs and SGXBs, the supergiant fast X--ray transients 
\citep[SFXTs, ][]{Smith2004:fast_transients,zand2005,Sguera2005,Negueruela2006:ESASP604}. 
SFXTs are associated with OB supergiant stars but, unlike SGXBs, 
show the most dramatic manifestation of their activity as bright outbursts 
during which they experience an increase in X--ray luminosity by up to a factor of $10^{5}$,
reaching peak luminosities of 10$^{36}$--10$^{37}$~erg s$^{-1}$. 
These bright outbursts last a few hours in the hard X--ray band \citep[][]{Sguera2005,Negueruela2006} and,  
although the outbursts can last up to a few days in the soft X--ray band
\citep[e.g.][]{Romano2007,Romano2013:Cospar12}, they are still 
significantly shorter than those of typical BeXBs. 
The hard X--ray spectra, qualitatively similar to those of HMXBs 
that host accreting neutron stars (NS), are generally modelled with 
often heavily absorbed power laws with a high energy cut-off. 
Therefore, it is tempting to assume that all SFXTs host a neutron star, 
even if pulse periods have only been measured for only a few systems. 
Currently the SFXT class consists of 14 objects 
\citep[see][and references therein]{Romano2014:sfxts_catI} 
and as many candidates (transients showing an SFXT behaviour but still lacking 
optical identification with an OB supergiant companion). 

The physical mechanisms causing the bright SFXT outbursts are still uncertain. 
In the last decade several models have been proposed that can be divided in two main
groups, related to either the properties of the wind from the supergiant companion 
\citep{zand2005,Walter2007,Negueruela2008,Sidoli2007} 
or the properties of the compact object, in particular the presence of mechanisms regulating or inhibiting accretion
(the propeller effect, \citealt[][]{Grebenev2007,grebenev09}; magnetic gating, \citealt[][]{Bozzo2008}).
A model of quasi-spherical accretion onto neutron stars involving      
hot shells of accreted material above the magnetosphere 
\citep[][and references therein]{Shakura2014:bright_flares}, has recently been proposed. 

%
%
The transient IGR~J17544$-$2619 is the prototypical SFXT.  
It was discovered by \inte\ on 2003 September 17 \citep{Sunyaev2003} 
during a 2-hr flare that reached an 18--25~keV flux of $6\times10^{-10}$~\ferg\ (160\,mCrab). This source was later observed 
in very bright states, lasting up to 10 hours, with 20--40 keV fluxes up to $6\times10^{-10}$~\ferg\ 
\citep[400~mCrab;][]{Grebenev2003:17544-2619,Grebenev2004:17544-2619,Sguera2006,Walter2007,Kuulkers2007}. 
Some flares were also found in archival \sax\ data \citep{zand2004:17544bepposax}.  
Several outbursts were also observed by \sw\ 
\citep{Krimm2007:ATel1265,Sidoli2009:sfxts_paperIII,Sidoli2009:sfxts_paperIV,
rlv+11,Romano2011:sfxts_paperVII,Farinelli2012:sfxts_paperVIII} 
and \suzaku, which caught a $\gtrsim$ day long outburst \citep[][]{Rampy2009:suzaku17544}.  

IGR~J17544$-$2619 is now a quite well studied binary.  
The primary is an O9Ib star \citep{Pellizza2006} at 3.6\,kpc \mbox{\citep{Rahoui2008}}, 
and the orbital period is $4.926\pm0.001$\,d  \citep[][]{Clark2009:17544-2619period,smith14}. 
While \citet[][]{Drave2012:17544_2619_pulsation} reported pulsations at $71.49\pm0.02$\,s 
from the region around the source that they attributed to a spin period, \citet{Drave2014:17544} did not confirm this detection.

\igrheavy\ is characterized by high variability. 
It was the first SFXT observed in detail during quiescence 
(at $L\sim5\times10^{32}$ erg\,s$^{-1}$). A {\it Chandra} observation 
\citep[][]{zand2005} showed that the the source is characterized by a very soft 
($\Gamma=5.9\pm1.2$) spectrum. Furthermore, this 
state of quiescence was followed by a bright flare, thus implying a 
dynamical range of at least 4 orders of magnitude. 
These observations, with their extreme luminosity changes occurring on such short time scales,
were interpreted in terms of accretion onto a compact object (probably a neutron star) 
from an inhomogeneous, or `clumpy', wind from the supergiant companion 
\citep[][]{zand2005}. 
Alternatively, \citet[][]{Bozzo2008} explained the large luminosity swings 
observed on time scales as short as hours in terms of transitions across the 
magnetic barriers. In this scenario, SFXTs 
with large dynamic range and $P_{\rm spin} \gtrsim 1000$\,s must have 
magnetar-like fields ($B\gtrsim 10^{14}$\,G). 

In this paper, we present the first firm detection of a cyclotron line in the
spectrum of an SFXT and hence the first direct measurement of its magnetic field. 

\section{Observations and analysis}\label{sec:obs}

\igrheavy\ was observed by \nustar\ on 2013~June~18--19, and near-simultaneously by \sw\ (Table~\ref{tab:obsid}). These observations were planned near orbital phase 0~\citep{smith14} to maximize a chance of detecting a flare.

%

\begin{table}
\caption{Observations of \igrheavy \label{tab:obsid}}
\begin{tabular}{rll}

\hline
\multicolumn{3}{c}{\nustar}\\
\hline
OBSID & 30002003002 & 30002003003\\
Start Date & 2013-06-18T22:16:07 & 2013-06-19T09:31:07\\
End Date & 2013-06-19T09:31:07 & 2013-06-19T23:41:07\\
Start MJD & 56461.9344668 & 56462.4059946\\
Exposure FPMA & 17533.22~s & 26238.50~s\\
Exposure FPMB & 17576.65~s & 26878.83~s\\
\hline
\multicolumn{3}{c}{\sw/XRT}	\\
\hline
OBSID & 00080201001 & 00080201003\\
Start Date & 2013-06-18T23:00:31 & 2013-06-19T00:42:08\\
End Date   & 2013-06-18T23:20:55 & 2013-06-19T00:56:55\\
  Start MJD & 56461.9587016  & 56462.0292600 \\ 
  Exposure & 1208.52~s & 885.08~s \\
\hline
\end{tabular}
\end{table}

\nustar\ data were extracted and reduced with \software{nustardas} v1.2.0 (14~June~2013), and 
\software{Heasoft~6.14}. We extracted events from a 40\arcsec\ radius circular region centred on the source. Background was extracted from a large source--free region on the same detector. Appropriate response matrices and ancillary response files for this observation were generated using \texttt{numkrmf} and \texttt{numkarf} respectively. We used \nustar\ responses from CALDB version 20130509. \nustar\ consists of two co-aligned telescopes, each with a focal plane module (FPMA and FPMB). In FPMB, the source position was strongly contaminated by stray light of nearby bright sources during OBSID~30002003002. \igrheavy\ showed flaring activity during this observation (Section~\ref{sec:flare}). 

%
The \sw/\xrt\ data were processed with standard procedures 
(\software{xrtpipeline} v0.12.8), filtering and screening criteria using 
\software{FTOOLS} (v6.15.1). 
Source events were accumulated within a circular region
with a radius of 20 pixels (1 pixel $\sim2\farcs36$).
Background events were accumulated from an annular source-free region
centred on IGR~J17544$-$2619 (inner/outer radii of 70/100 pixels). 
For our spectral analysis, ancillary response files were generated with \software{xrtmkarf}
to account for different extraction regions, vignetting, and PSF corrections.
We used the latest \xrt\ spectral redistribution matrices in CALDB (20140120).

Data were analysed in \software{Xspec} (v12.8.1). We used \sw/\xrt\ data from 0.3--10~keV and \nustar\ data in the energy range 3--50~keV. Data were grouped to have at least 20 source+background photons per bin, and $\chi^2$ statistics were used for fitting. We used atomic cross sections from \citet{vfk+96} and elemental abundances from \citet{wam00}.

\section{Timing}\label{sec:timing}

\begin{figure}
  \centering
  \includegraphics[width=0.5\textwidth]{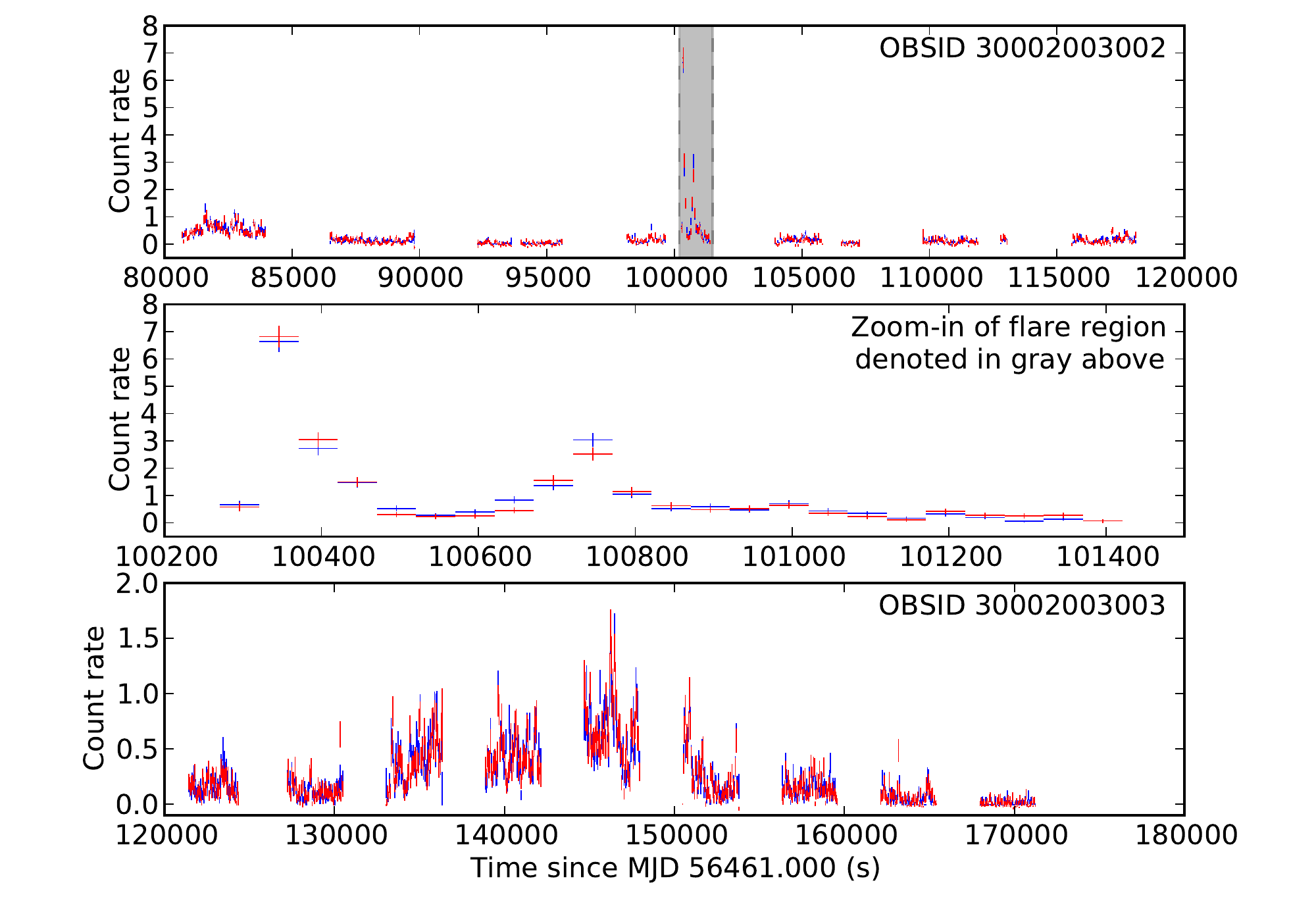}
  \caption{Background-subtracted \nustar\ lightcurves of \igrheavy\ OBSID~30002003002 (top panel) and OBSID~30002003003 (bottom panel). The middle panel zooms in on the flare region from OBSID~30002003002. Blue and red plus signs show count rates in 50~s bins for the focal plane modules FPMA and FPMB, respectively. For all panels, the X--axis is time since MJD~56461.0, Y--axes show counts~s$^{-1}$ in the 3--50~keV band. For reference, the average flux in OBSID~30002003003 in the entire 3--50~keV band is $(3.53\pm0.05) \times 10^{-11}$~\ferg\ (1~mCrab).}\label{fig:lc}
\end{figure}

Figure~\ref{fig:lc} shows the background-subtracted lightcurves with 50~s bins for both FPMs for the entire observation. OBSID 30002003002 shows strong flaring activity from \igrheavy, with a bright flare that is about ten times stronger than the average flux level (Section~\ref{sec:flare}). The source is less variable in OBSID 30002003003, with a dynamic range of just a factor of two. The average absorbed source flux in this OBSID is $(1.11\pm0.01) \times 10^{-11}$~\ferg\ in the $3-10$~keV band, consistent with the average unabsorbed source flux of $10^{-11}$~\ferg\ measured by \sw/\xrt\ in the $2-10$~keV band~\citep{rlv+11}. The total absorbed flux observed by \nustar\ is $(3.53\pm0.05) \times 10^{-11}$~\ferg in the $3-50$~keV band.

We searched the \nustar\ data for any pulsations in \igrheavy. No strong peaks are seen in the power spectra. 
An epoch folding search does not yield any strong periodicity either. In particular, we do not detect the claimed $71.49\pm0.02$\,s pulsation~\citep{Drave2012:17544_2619_pulsation}.
Further, we computed a power spectrum and renormalized it relative to the local mean power in order to search for statistically significant periodic signals. We found periodic signals at about 1455~s and 1940~s, which are integer fractions of the spacecraft's orbital period. The instrumental origin was confirmed when we extracted photons from background regions far from the source, and found peaks at the same periods.
 We conclude that \igrheavy\ does not show any strong pulsations in the range of 1~s to about 2000~s, consistent with \citet{Drave2014:17544}.

\section{Flare}\label{sec:flare}

\begin{figure}
  \centering
  \includegraphics[width=0.5\textwidth]{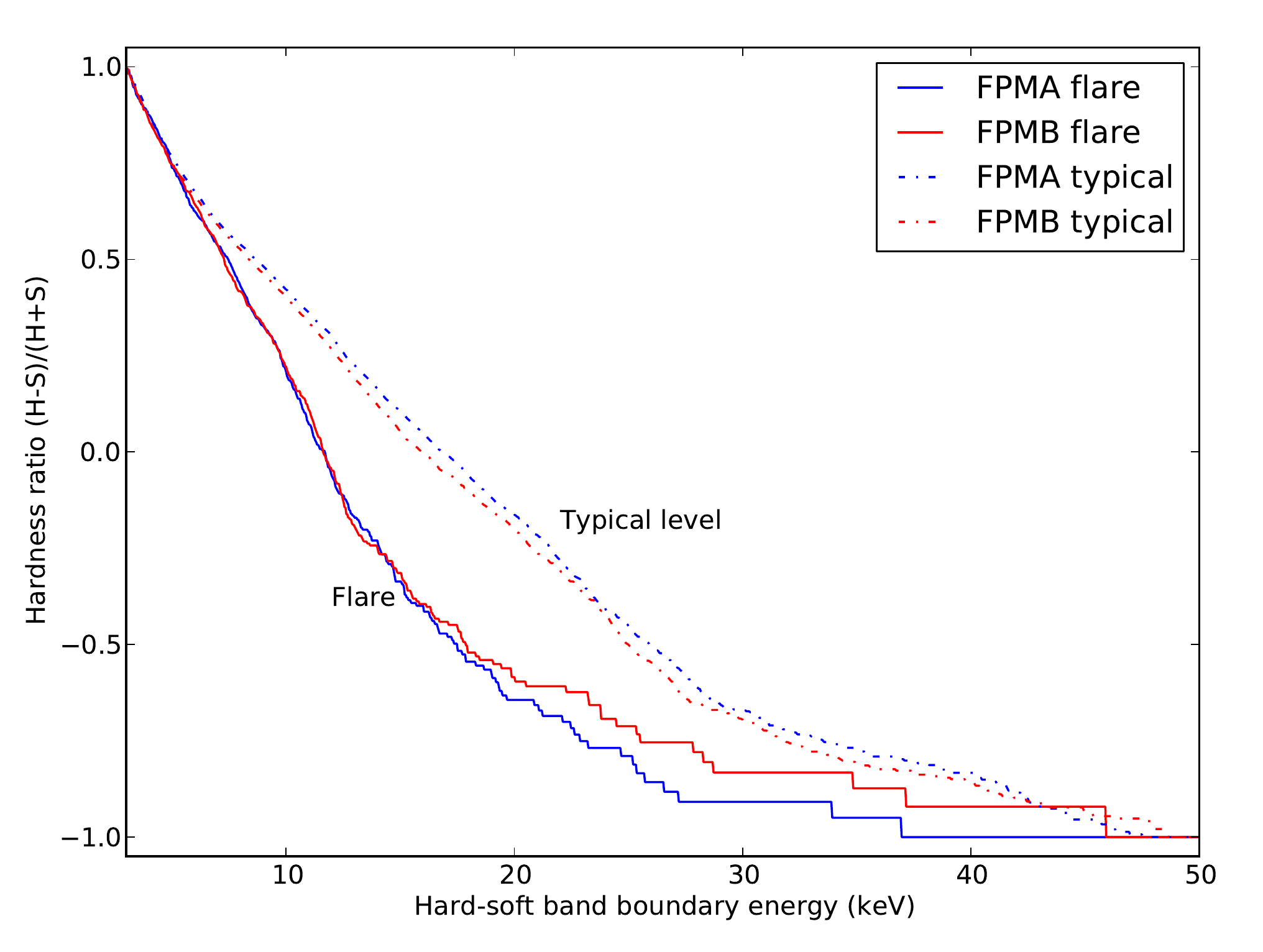}
  \caption{Hardness ratio of \igrheavy\ as a function of energy, during the flare and in quiescence. For each energy $E$, we define the $3-E$~keV band as the soft band, and $E-50$~keV band as the hard band. The $Y$-axis shows the hardness ratio, defined as (H-S)/(H+S). Red curves are for FPMA, and blue curves are for FPMB. The solid lines are cumulative fluxes of the flare (Figure~\ref{fig:lc}), compared with quiescent fluxes from OBSID~30002003003. The flare is softer than the ``typical'' state, the difference being most prominent at 15--20~keV. For example, with 3--15~keV and 15--50~keV bands, the hardness ratio is about 0.1 in the typical state, but falls to about $-0.3$ during the flare.}\label{fig:flarehard2}
\end{figure}

\igrheavy\ is known for strong flaring behaviour. \nustar\ detected a flare during OBSID~30002003002, starting approximately at MJD~56462.161 and spanning about 220~seconds (Figure~\ref{fig:lc}, middle panel). It was followed by a smaller flare about 400~s later. The spectrum of the first flare is relatively flat from $3-10$~keV and falls off at higher energies. We calculate the model--independent flux for the source and the flare using \nustar\ response files. The average absorbed flux in the flare is $(3.1\pm0.1) \times 10^{-10}$~\ferg\ (9~mCrab) in the 3--50~keV range, about an order of magnitude higher than the average flux of $(3.54\pm0.05) \times 10^{-11}$~\ferg\ (1~mCrab) measured in OBSID~30002003003. This is consistent with typical flares observed near periastron from this source ~\citep{rlv+11}. The source becomes softer during the flare (Figure~\ref{fig:flarehard2}). The average absorbed flux of the second flare is $(1.5\pm0.1) \times 10^{-10}$~\ferg.

Broadband ($\sim$0.2--60\,keV) flare spectra ($\sim10^{-9}$~\ferg)
are typically modelled as an absorbed cut-off power--law or an
absorbed power--law with an exponential cut-off. For example,
\citet[][]{Rampy2009:suzaku17544} fit the \suzaku\ XIS+PIN data on the 2008 March 31
outburst with an  absorbed power--law with an exponential cut-off with
$\Gamma = 0.9$, and $E_{\rm fold} = 10.5$~keV;
\citet[][]{rlv+11} adopt an absorbed power--law with a high energy cut-off for the \sw\ \textit{BAT}+\xrt\ data on the
2009 June 6 outburst and find $\Gamma = 0.6$, $E_{\rm cut} = 3.3$~keV, and $E_{\rm fold} = 8.1$~keV. 
However, our flare data are not fit well by a simple absorbed blackbody or absorbed cut--off power--law model, which give $\chi^2_\nu = 1.76$ and 1.4 with 47 and 46 degrees of freedom respectively. The simplest model for the flare spectrum is an unabsorbed power--law with two breaks (\texttt{bkn2pow)} at 8.9 and 11.1~keV. For this model, we get $\chi^2_\nu = 0.93$ with 44 degrees of freedom. 

\begin{figure*}
  \centering
  \includegraphics[width=0.9\textwidth]{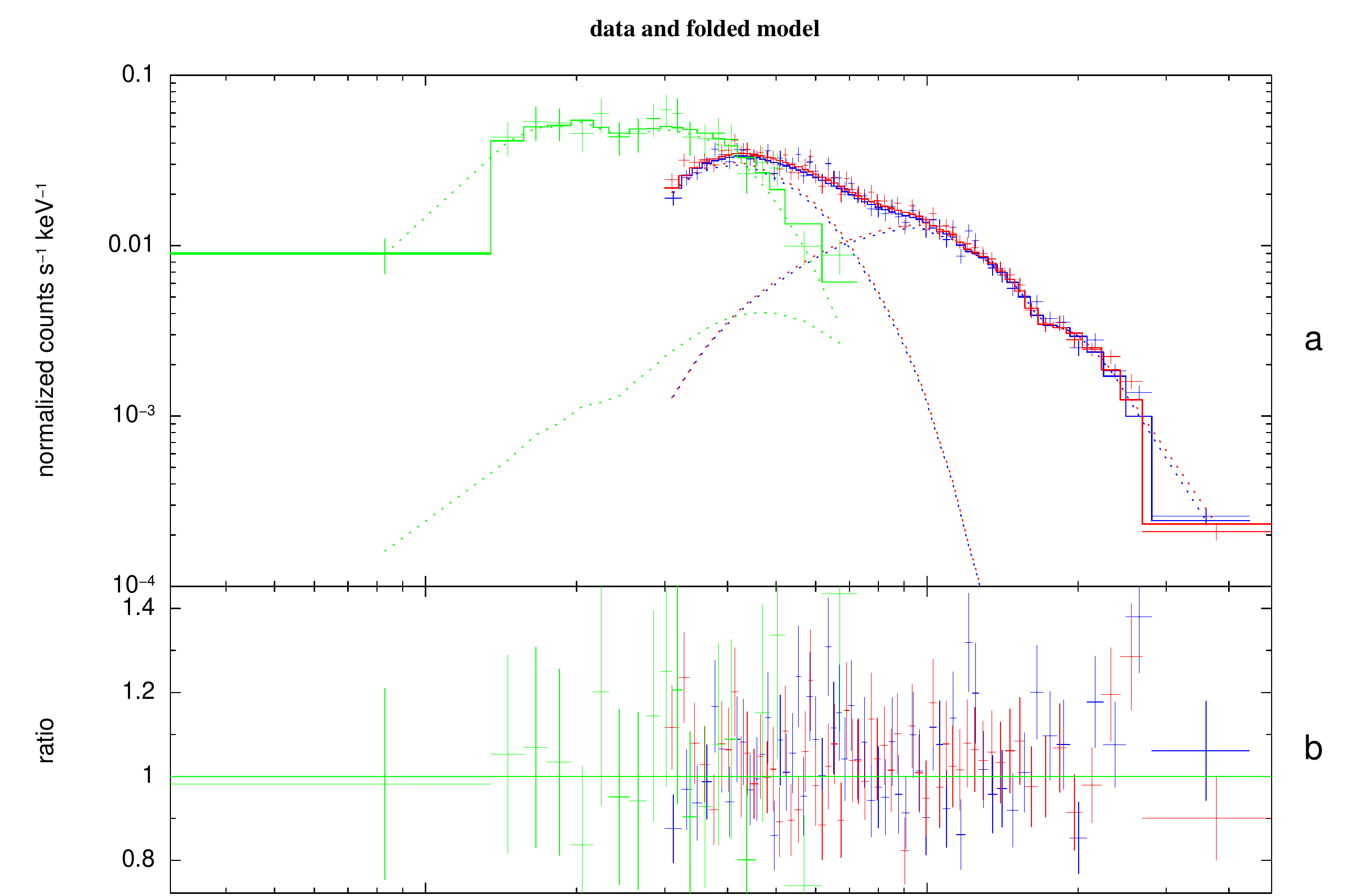}
  \includegraphics[width=0.9\textwidth]{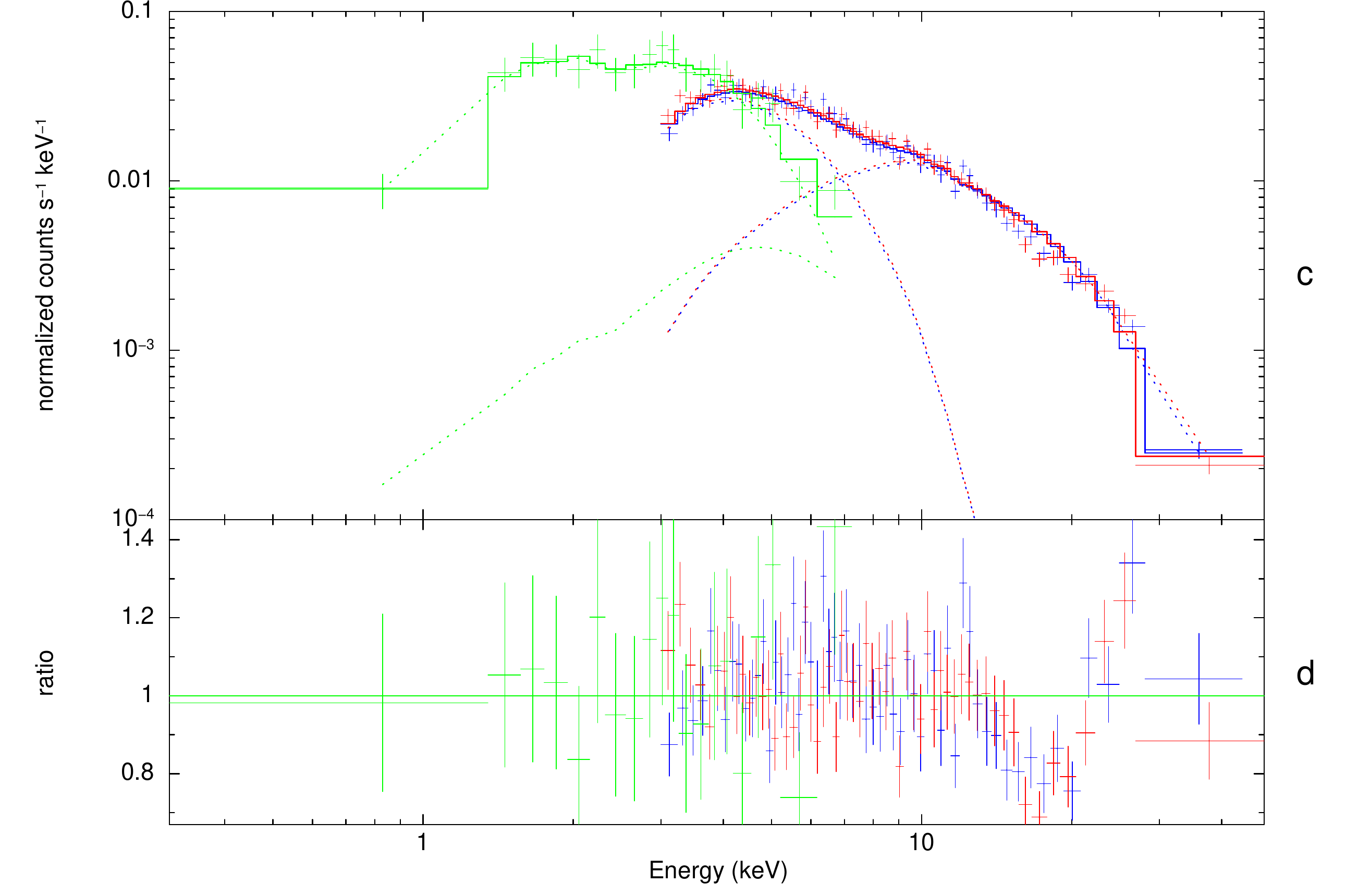}
  \caption{Joint fit to \nustar\ and \sw/\xrt\ data with \texttt{bbodyrad + nthcomp} as the continuum model. Blue, red and green symbols denote data from \nustar\ FPMA, \nustar\ FPMB and \sw/\xrt, respectively. For plotting \nustar\ data have been re-binned to a minimum SNR 10 in each bin--actual fitting was done with smaller bins with at least 20 photons each for both: \nustar\ and \sw/\xrt. We allow a scaling factor between \nustar\ FPMA, FPMB and \sw/\xrt\ fluxes. Panel a shows the best-fit with the continuum and a single cyclotron line (no harmonics). The ratio data to the model (Panel b) is relatively flat, as expected for a well--fit model. Panel c shows the same model with the cyclotron line deleted (but without refitting). The ratio of data to the model (panel d) clearly show the cyclotron line.}\label{fig:spec3}
\end{figure*}


\begin{table*}
\begin{minipage}{126mm}
\caption{Spectral fits for \igrheavy\ with continuum model \texttt{I} (\texttt{bbodyrad + nthcomp}) \label{tab:cyclotron}}
\begin{tabular}{clr@{ }lr@{ }lr@{ }lr@{ }l}
\hline
Model & Parameter & \multicolumn{8}{c}{Model} \\
component & name & \multicolumn{2}{c}{No line} & \multicolumn{2}{c}{Single line} & \multicolumn{2}{c}{Line + harmonic} & \multicolumn{2}{c}{Two lines}\\
\hline
\multicolumn{10}{c}{\textit{Continuum parameters}}\\
\textit{$N_{\rm H}$} & (10$^{22}~{\rm cm}^{-2}$) & 1.7 & $\pm$0.3 & 1.52 & $_{-0.23}^{+0.28}$ & 1.43 & $_{-0.23}^{+0.28}$ & 1.45 & $_{-0.23}^{+0.28}$ \\
%
\multirow{2}{*}{\textit{bbodyrad}} & \textit{kT} (keV) & 0.95 & $\pm$0.04 & 1.04 & $\pm$0.02 & 1.07 & $\pm$0.03 & 1.06 & $_{-0.03}^{+0.04}$ \\
& \textit{norm} & 1.06 & $_{-0.14}^{+0.17}$ & 0.92 & $_{-0.08}^{+0.10}$ & 0.83 & $_{-0.08}^{+0.11}$ & 0.85 & $_{-0.06}^{+0.10}$ \\
%
\multirow{3}{*}{\textit{nthcomp}}& $\Gamma$$^{\rm a}$ & 1.21 & $_{-0.05}^{+0.04}$ & 1.00 & $_{-***}^{+0.03}$ & 1.00 & $_{-***}^{+0.06}$ & 1.00 & $_{-***}^{+0.05}$ \\
& \textit{kT$_e$} (keV) & 5.66 & $_{-0.23}^{+0.24}$ & 5.04 & $\pm$0.08 & 6.4 & $_{-0.9}^{+6.2}$ & 5.8 & $_{-0.5}^{+5.5}$ \\
& \textit{norm ($10^{-6}$)} & 94 & $_{-27}^{+29}$ & 4.0 & $_{-0.2}^{+18}$ & 3.1 & $_{-0.5}^{+50}$ & 3.4 & $_{-0.8}^{+64}$ \\
%
\multirow{2}{*}{\textit{X-norm constant}} &\textit{FPMB} & 1.12 & $\pm$0.02 & 1.12 & $\pm$0.02 & 1.12 & $\pm$0.02 & 1.12 & $\pm$0.02 \\
&\textit{\sw/XRT} & 1.33 & $_{-0.08}^{+0.09}$ & 1.34 & $_{-0.08}^{+0.09}$ & 1.35 & $_{-0.08}^{+0.09}$ & 1.35 & $_{-0.08}^{+0.09}$ \\
%
\multicolumn{10}{c}{\textit{Cyclotron lines}}\\
\multirow{3}{*}{\textit{Line 1}} & \textit{Energy} (keV) & \multicolumn{2}{c}{\nodata} & 16.9 & $\pm$0.3 & 16.9 & $\pm$0.3 & 17.0 & $\pm$0.3 \\
& \textit{Width} (keV) & \multicolumn{2}{c}{\nodata} & 1.6 & $_{-0.5}^{+0.6}$ & 3.5 & $_{-1.0}^{+0.8}$ & 3.0 & $_{-0.7}^{+0.8}$ \\
& \textit{Depth} (keV) & \multicolumn{2}{c}{\nodata} & 0.40 & $_{-0.06}^{+0.07}$ & 0.58 & $_{-0.14}^{+0.06}$ & 0.53 & $_{-0.09}^{+0.06}$ \\
%
\multirow{3}{*}{\textit{Line 2}} & \textit{Energy} (keV) & \multicolumn{2}{c}{\nodata} & \multicolumn{2}{c}{\nodata} & (33.8)&$^{\rm b}$ & 32.9 & $_{-1.1}^{+1.3}$ \\
& \textit{Width} (keV) & \multicolumn{2}{c}{\nodata} & \multicolumn{2}{c}{\nodata} & 9.8 & $\pm$5.0 & 6.6 & $_{-2.0}^{+6.6}$ \\
& \textit{Depth} (keV) & \multicolumn{2}{c}{\nodata} & \multicolumn{2}{c}{\nodata} & 1.2 & $_{-0.6}^{+1.4}$ & 0.9 & $_{-0.4}^{+0.3}$ \\
%
\multicolumn{10}{c}{\textit{Quality of fit}}\\
\textit{Degrees of freedom} & & \multicolumn{2}{c}{503} & \multicolumn{2}{c}{500} & \multicolumn{2}{c}{498} & \multicolumn{2}{c}{497} \\
\textit{$\chi^2$} & & \multicolumn{2}{c}{515.9} & \multicolumn{2}{c}{477.2} & \multicolumn{2}{c}{467.7} & \multicolumn{2}{c}{467.1} \\
\textit{$\Delta \chi^2$} & & \multicolumn{2}{c}{\nodata} & \multicolumn{2}{c}{$-$38.7} & \multicolumn{2}{c}{$-$48.2} & \multicolumn{2}{c}{$-$48.8}\\
\hline
\end{tabular}
\medskip\\
We allow relative scaling of \nustar\ FPMA, FPMB and \sw/XRT data. The best fit values for the cross-normalization (X-norm) constants are included in the table.\\
$^{\rm a}$In fits including the cyclotron lines, $\Gamma$ gets pegged at its lower limit of 1.0. Hence we give only one--sided error bars on this parameter.\\
$^{\rm b}$Energy of the harmonic is defined as two times the energy of the fundamental, and is not a free parameter.

\end{minipage}
\end{table*}

\begin{table*}
\begin{minipage}{126mm}

\caption{Spectral fits for \igrheavy\ with continuum model \texttt{II} (\texttt{bbodyrad + cutoffpl}) \label{tab:cyclotron2}}

\begin{tabular}{clr@{ }lr@{ }lr@{ }lr@{ }l}
\hline
Model & Parameter & \multicolumn{8}{c}{Model} \\
component & name & \multicolumn{2}{c}{No line} & \multicolumn{2}{c}{Single line} & \multicolumn{2}{c}{Line + harmonic} & \multicolumn{2}{c}{Two lines}\\
\hline
\multicolumn{10}{c}{\textit{Continuum parameters}}\\
\textit{$N_{\rm H}$} & (10$^{22}~{\rm cm}^{-2}$) & 1.6 & $\pm$ 0.3 & 1.38 & $_{-0.22}^{+0.26}$ & 1.34 & $_{-0.22}^{+0.18}$ & 1.4 & $\pm$ 0.2 \\
%
\multirow{2}{*}{\textit{bbodyrad}} & \textit{kT} (keV) & 0.99 & $_{-0.04}^{+0.03}$ & 1.097 & $_{-0.006}^{+0.02}$ & 1.115 & $_{0.006}^{+0.03}$ & 1.102 & $_{-0.006}^{+0.02}$ \\
& \textit{norm} & 1.04 & $_{-0.12}^{+0.17}$ & 0.78 & $_{-0.06}^{+0.08}$ & 0.74 & $\pm$ 0.08 & 0.76 & $_{-0.06}^{+0.07}$ \\
%
\multirow{3}{*}{\textit{cutoffpl}} & \textit{$\Gamma$$^{\rm a}$} & -1.1 & $_{-0.2}^{+0.3}$ & -3.0 & $_{-***}^{+0.4}$ & -2.8 & $_{-***}^{+1.7}$ & -3.0 & $_{-***}^{+0.4}$ \\
& \textit{E$_{cut}$} (keV) & 6.6 & $_{-0.5}^{+0.7}$ & 4.04 & $_{-0.05}^{+0.02}$ & 4.75 & $_{-0.5}^{+0.03}$ & 4.18 & $_{-0.08}^{+0.02}$ \\
& \textit{norm ($10^{-6}$)} & 21 & $_{-8}^{+13}$ & 0.72 & $_{-0.04}^{+0.01}$ & 0.54 & $_{-0.14}^{+0.02}$ & 0.67 & $\pm$ 0.02 \\
%
\multirow{2}{*}{\textit{X-norm constant}} &\textit{FPMB} & 1.12 & $\pm$0.02 & 1.12 & $\pm$0.02 & 1.12 & $\pm$0.02 & 1.12 & $\pm$0.02 \\
&\textit{\sw/XRT} & 1.34 & $_{-0.08}^{+0.09}$ & 1.35 & $_{-0.08}^{+0.09}$ & 1.35 & $_{-0.07}^{+0.08}$ & 1.35 & $\pm$ 0.08 \\
%
\multicolumn{10}{c}{\textit{Cyclotron lines}}\\
%
\multirow{3}{*}{\textit{Line 1}} & \textit{Energy} (keV) & \multicolumn{2}{c}{\nodata} & 16.8 & $\pm$ 0.3 & 16.6 & $_{-0.3}^{+0.2}$ & 16.9 & $_{-0.3}^{+0.2}$ \\
& \textit{Width} (keV) & \multicolumn{2}{c}{\nodata} & 2.6 & $_{-0.3}^{+0.6}$ & 4.6 & $_{-0.3}^{+0.8}$ & 3.1 & $_{-0.3}^{+0.6}$ \\
& \textit{Depth} (keV) & \multicolumn{2}{c}{\nodata} & 0.49 & $_{-0.04}^{+0.06}$ & 0.72 & $_{-0.03}^{+0.18}$ & 0.54 & $_{-0.04}^{+0.20}$ \\
%
\multirow{3}{*}{\textit{Line 2}} & \textit{Energy} (keV) & \multicolumn{2}{c}{\nodata} & \multicolumn{2}{c}{\nodata} & (33.2) & $^{\rm b}$ & 30.0 & $_{-0.5}^{+1.9}$ \\
& \textit{Width} (keV) & \multicolumn{2}{c}{\nodata} & \multicolumn{2}{c}{\nodata} & 7.4 & $_{-3.5}^{+4.3}$ & 1 & $_{-***}^{+7}$~$^{\rm c}$ \\
& \textit{Depth} (keV) & \multicolumn{2}{c}{\nodata} & \multicolumn{2}{c}{\nodata} & 1.09 & $_{-0.10}^{+0.09}$ & 0.7 & $_{-0.3}^{+0.2}$ \\
%
\multicolumn{10}{c}{\textit{Quality of fit}}\\
\textit{Degrees of freedom} & & \multicolumn{2}{c}{503} & \multicolumn{2}{c}{500} & \multicolumn{2}{c}{498} & \multicolumn{2}{c}{497} \\
\textit{$\chi^2$} & & \multicolumn{2}{c}{516.0} & \multicolumn{2}{c}{473.4} & \multicolumn{2}{c}{467.4} & \multicolumn{2}{c}{465.3} \\
\textit{$\Delta \chi^2$} & & \multicolumn{2}{c}{0.0} & \multicolumn{2}{c}{$-$42.6} & \multicolumn{2}{c}{$-$48.6} & \multicolumn{2}{c}{$-$50.7}\\
\hline
\end{tabular}
\medskip \\
We allow relative scaling of \nustar\ FPMA, FPMB and \sw/XRT data. The best fit values for the cross-normalization (X-norm) constants are included in the table.\\
$^{\rm a}$In fits including the cyclotron lines, $\Gamma$ gets pegged at its lower limit of $-3.0$. Hence we give only one--sided error bars on this parameter.\\
$^{\rm b}$Energy of the harmonic is defined as two times the energy of the fundamental, and is not a free parameter.\\
$^{\rm c}$The minimum width of line 2 gets pegged at its lower limit of 1~keV before obtaining $\Delta\chi^2=1.0$, so we do not give a lower limit.

\end{minipage}
\end{table*}


\section{Spectrum}\label{sec:spectrum}

For spectral modelling, we only use data from OBSID~30002003003, where the source is in a steady state. We used \nustar\ data extracted with a 40\arcsec\ extraction region, grouped to make bins of at least 20 photons and \sw/\xrt\ data from both \sw\ observations. The spectrum can be fit by a two-component model consisting of a $\sim1$~keV blackbody and a harder, non--thermal component (Figure~\ref{fig:spec3}). 
This non-thermal component can be interpreted as a Comptonized spectrum with seed photons from the blackbody -- indeed, a non--thermal Comptonization model (\texttt{nthcomp}) with $\Gamma = 1.2$ and $kT_e = 5.7$~keV gives a reasonable fit (Table~\ref{tab:cyclotron}). 
Alternately, this component is also fit well by the empirical cut-off power--law model with $\Gamma = -1.1$ and $E_{\rm cut} = 6.7$~keV (Table~\ref{tab:cyclotron2}). Hereafter we refer to these as continuum models \texttt{I} and \texttt{II} respectively.

We can calculate the size of the emitting area of the blackbody component from its normalization ($norm$\footnote{\url{http://heasarc.gsfc.nasa.gov/xanadu/ xspec/manual/ XSmodelBbodyrad.html}.} in \textit{Xspec}) and distance to the object: $norm$ = $R_{\rm km}^2 / {D}_{10}^2$. Using a nominal distance of 3.6~kpc to \igrheavy~\citep{rcl+08} and assuming a circular emitting area, the best-fit $norm$ values correspond to a radius $R\approx0.3$~km. This is consistent with the size of an accretion hotspot on the NS for low accretion rates~\citep{fkr02}.

Regardless of the continuum model, the fits show systematic residuals mimicking absorption features. Good fits can be obtained only on introducing cyclotron absorption features in the model (Figure~\ref{fig:spec3}). We tested the 
presence and significance of these lines with various extraction apertures and binning methods.
Further, we also tested the presence of a harmonic in two ways: enabling the harmonic in \texttt{cyclabs}, and adding an independent line at higher energy. All these tests gave consistent results: the spectral fits are significantly better when a cyclotron line is included in the spectral model. The fits improve further when the cyclotron line harmonic is also added in the fit. Adding an independent higher energy line gives results broadly consistent with the location of a harmonic. 



In continuum model~\texttt{I}, adding a cyclotron line gives $\Delta \chi^2 = 38.7$ for three more degrees of freedom. The best--fit line energy is $E_{\rm cyc} = 16.9\pm0.3$~keV (Table~\ref{tab:cyclotron}). For continuum model~\texttt{II}, adding a cyclotron line gives $\Delta \chi^2 = 42.6$ for three more degrees of freedom (Table~\ref{tab:cyclotron2}). The best--fit line energy, $E_{\rm cyc} = 16.8\pm0.3$~keV, agrees with the fit for model~\texttt{I}. In both cases, adding a harmonic decreases the $\chi^2$ further. If we introduce a second, independent absorption feature, its best--fit energy agrees with the expected harmonic to within 1-$\sigma$ for continuum model \texttt{I}. For continuum model \texttt{II}, the best--fit energy of this absorption feature is slightly lower than twice the fundamental. This slight difference in energies is seen in other X-ray binaries as well~\citep{cw12}.

\begin{figure}
  \centering
  \includegraphics[trim=0.4cm 0cm 0cm 0.5cm,clip=true,width=0.45\textwidth]{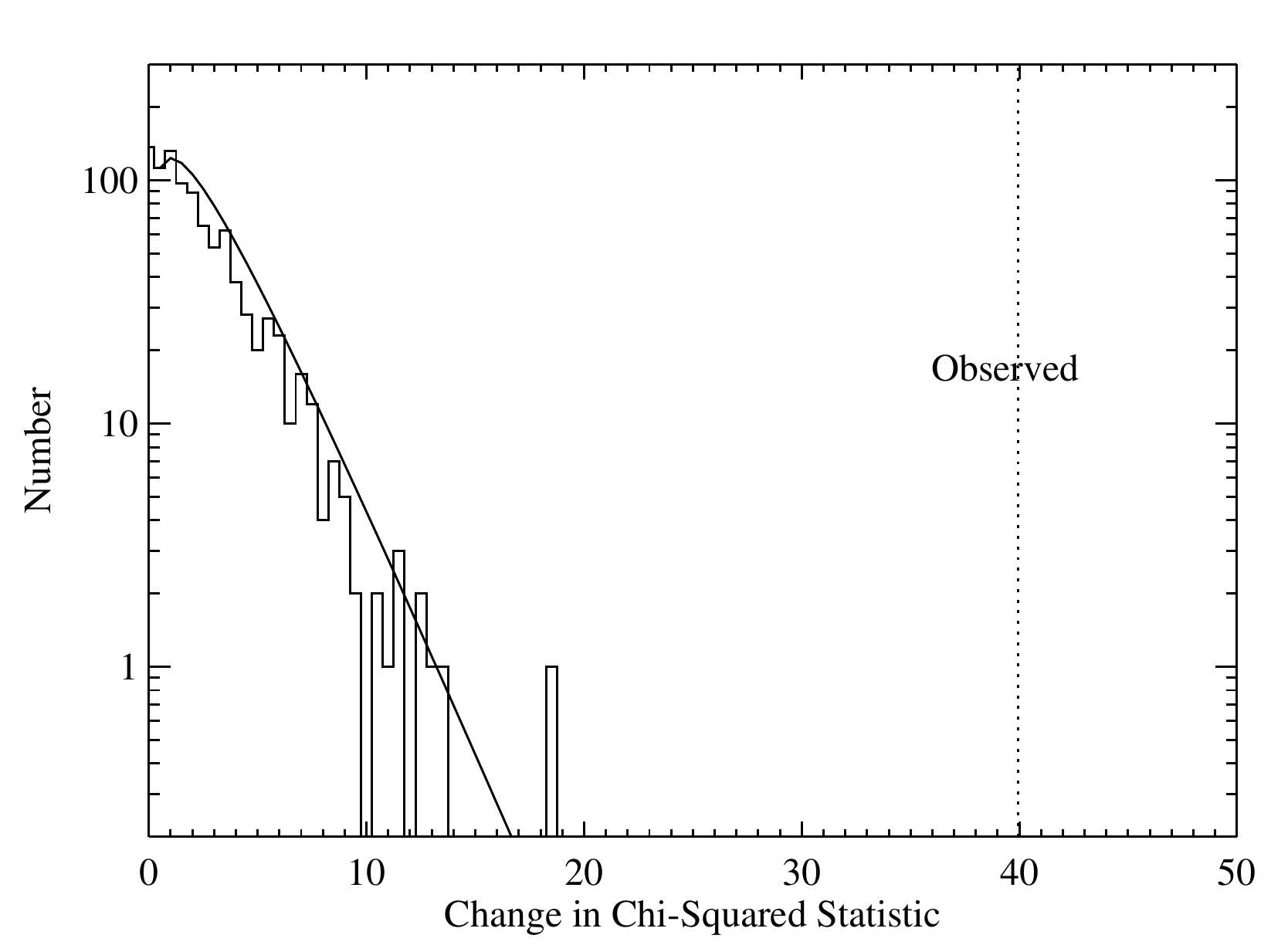}
  \caption{Results of Monte-Carlo simulations for testing line significance for continuum model \texttt{II} (Section~\ref{sec:spectrum}). We simulated one thousand fake \nustar\ spectra for the continuum--only model (\texttt{bbodyrad + cut-offpl}), and attempted to fit them with a continuum+line model with $N_H$ fixed at the value obtained with a joint \sw/\xrt\ fit. The solid histogram shows $\Delta \chi^2$ values obtained in the simulations, and the smooth curve is a $\chi^2$ distribution with three degrees of freedom. The $\Delta \chi^2$ attained in actual data (dashed vertical line) is significantly higher than values attained in simulations. This $\Delta \chi^2$ value differs that in Table~\ref{tab:cyclotron2} due to the simplifying assumptions made in these simulations (Section~\ref{sec:spectrum}). }\label{fig:ftest}
\end{figure}

We checked for the significance of the line depth using three methods for both continuum models. We consider the case with only the fundamental line without any harmonics. We allow the line depth to vary over a wide range, so as to search for cyclotron absorption or emission features. \textit{(i) F-test:} Based on the
improvement in $\chi^2$ by adding the line, we can calculate a false detection probability for the line\footnote{Note that the line depth was allowed to be positive as well as negative, so that the null model (no line) is not a boundary case for the F-test.}. 
For continuum model \texttt{I}, we get $p = 1.7\times 10^{-8}$ while for continuum model \texttt{II}, $p = 2.3\times10^{-9}$.
\textit{(ii) Non-zero line depth:} We considered models with the fundamental line 
only, and stepped through a grid of
values of the line depth and width with the \software{Xspec} command
\texttt{steppar}, and noted the change in $\chi^2$. For continuum model \texttt{I}, we 
find that changing the line depth to zero gives a minimum $\Delta \chi^2$ of 52, corresponding to a 
7-$\sigma$ detection. The constraints were even stronger for continuum model \texttt{II}.
\textit{(iii) Monte--Carlo simulations:} Further, we tested the line significance by simulating
spectra using the \software{Xspec} script \texttt{simftest}. We used the continuum
model~\texttt{II}, consisting of a blackbody and a cut--off powerlaw as our
\textit{null hypothesis}. We simulated fake spectra from this model and fit them
with (a) only continuum, and (b) continuum + cyclotron line.  To improve the
speed and convergence of the fits, we performed simulations using only the two
{\em NuSTAR} modules, fixing the column density to the value found when \xrt\ was
included.  We repeated this test 1000 times and noted the change in $\chi^2$
obtained by adding a cyclotron line of similar width (within the 90\% confidence
region obtained with the actual data). Since the cyclotron line adds three free
parameters, we expect that the histogram of $\Delta \chi^2$ values should follow
a $\chi^2$ distribution with three degrees of freedom. This is indeed the case,
as seen in Figure~\ref{fig:ftest}. The highest $\Delta \chi^2$ obtained in our
simulation is 18.7, significantly lower than $\Delta \chi^2 = 41.2$ obtained in
real data.We estimate that $10^7 - 10^8$ simulations would be required to get
$\Delta \chi^2 > 40$ in one of them. Performing such a large number of simulations 
is technically infeasible. However,
scaling from our 1000 simulations, we obtain a line significance of
$>$5-$\sigma$.  We repeated this test with continuum model~\texttt{I}.  The
observed $\Delta \chi^2$ for this model is 37.1, but the maximum value we obtain
in 1000 simulations is 12, confirming the high significance of the cyclotron
line.

We tested the presence of a line at 8.5~keV by adding a model component with half the energy and half the width as the 17~keV line. We do not detect any significant absorption near 8.5~keV, with 3-$\sigma$ limits on line depth as $D_{8.5}^{\mathtt{I}} < 0.19$ and $D_{8.5}^{\mathtt{II}} < 0.15$ for continuum models \texttt{I} and \texttt{II}, respectively.

\section{Discussion\label{sec:discuss}}
Despite a decade of investigation, the mechanisms responsible 
for the flaring behaviour of SFXTs are still far from certain. 
Several, non mutually exclusive, models have been proposed, depending on 
the donor star wind and/or the accreting neutron star properties. 
For the `clumpy wind' models \citep{zand2005,Walter2007,Negueruela2008,Sidoli2007} 
the common key parameters are the geometry and inhomogeneity 
of the stellar wind from the supergiant donor star. 
For the `gating' models, mechanisms are required to regulate or inhibit accretion 
(the propeller effect--see \citealt{Grebenev2007}; or magnetic gating--see \citealt{Bozzo2008}). 
In particular, in the magnetic gating model \citep[][and references therein]{Bozzo2008}, 
the large intensity swings observed in IGR~J17544$-$2619 are explained 
in terms of slowly rotating ($P_{\rm spin} \gtrsim 1000$\,s) magnetar ($B\gtrsim 10^{14}$\,G) 
and a switch on/off of the source due to the propeller effect.
The quasi-spherical accretion model \citep[][and references therein]{Shakura2014:bright_flares},
on the other hand, featuring hot shells of accreted gas above the magnetosphere
of a slowly rotating ($P_{\rm spin} \gtrsim 1000$\,s) neutron star, can be applied to the
bright fast flares of SFXTs even without invoking magnetar-like B fields, when
the mass accretion rate increases due to the sporadic capture of magnetized stellar
wind plasma.

It is clear that knowledge of the magnetic field is a powerful discriminator
among various models. 
Until now, however, no measurements of the cyclotron lines were available, 
hence the magnetic field could only be estimated indirectly from the empirical
relationship of \citet[][]{Coburn2002}, using the cut-off energy derived from 
spectral fitting as a proxy for the magnetic field ($B$). 
For SFXTs, the typical cut-off energies are at about 10--20 keV, so the estimated
magnetic fields range from $2\times10^{12}$\,G for XTE~J1739.1$-$302, 
to about (2--$3)\times10^{12}$\,G  for IGR J17544$-$2619
\citep[][]{Sidoli2009:sfxts_paperIII}, and 
$\la 3\times10^{12}$\,G for AX~J1841.0$-$0536 \citep{Romano2011:sfxts_paperVII}. 

Detections of cyclotron features in HMXBs are still scarce, with under 20 confirmed detections before \nustar~\citep{cw12}.
Among those is the cyclotron line at $33 \pm 4$~keV reported in the \textit{candidate} 
SFXT IGR~J16493$-$4348 \citep[][]{dcl+11}, implying $B \approx 4 \times 10^{12}$~G. 
Our NuSTAR spectrum provides the very first measurement of such a feature in a confirmed 
SFXT, the prototype of the class IGR J17544$-$2619, at $16.8\pm0.3$~keV. 
Our data also shows hints of a line harmonic at an energy consistent with twice the fundamental, though slightly lower values are preferred (Tables~\ref{tab:cyclotron}, \ref{tab:cyclotron2}). The observed energy of cyclotron lines depends on the local magnetic field and the gravitational redshift caused by the neutron star:
\begin{equation}
B_{12} = \frac{E_{\rm cyc}}{11.6{\rm~keV}} (1 + z)
\end{equation}
where $B_{12}$ is the magnetic field in units of $10^{12}$~G. Hence, we conclude that the compact object in \igrheavy\ is indeed a neutron star, with magnetic field strength $B = (1.45\pm0.03) \times 10^{12}~G (1 + z)$. The gravitational redshift factor $(1 + z)$ is typically in the range of 1.25--1.4 for neutron stars~\citep{cw12}, but may be a bit higher for \igrheavy\ due to the higher mass of the neutron star~\citep{bhalerao12}. This $B$ value is consistent with the the $B \la 3\times10^{12}$~G constraint from spectral modelling~\citep[][]{Sidoli2009:sfxts_paperIII}.

An alternate interpretation is that the feature is a proton cyclotron line. In this latter case, the inferred magnetic field strength is $B_{12}^\prime = (m_{\rm p}/m_{\rm e}) B_{12}$ where $m_{\rm p}$ and $m_{\rm e}$ are proton and electron masses respectively.
This corresponds to a magnetic field strength $B^\prime = (2.66\pm0.06) \times 10^{15}~G (1 + z)$. In such fields, the equivalent width of lines is expected to be very low -- just few eV -- due to vacuum polarization (\citealt{hl03}, but also see \citealt{tem+13}). This contrasts strongly with the measured 2.2~keV equivalent width of the fundamental. Further, this latter magnetic field also contradicts the constraint from \citet{Sidoli2009:sfxts_paperIII}. As a result, we rule out this possibility that this absorption feature is a proton cyclotron line.

The energy of the cyclotron line, and the inferred magnetic field of \igrheavy\ is comparable to typical values measured in other X-ray binaries~\citep{cw12}. Furthermore, cyclotron line harmonics tend to have energies slightly lower than the corresponding multiple of the fundamental~\citep{cw12} -- as seen in our data, too.

Thus, the neutron star in IGR J17544$-$2619 is definitely not a magnetar, implying that one of 
the key requirements of the magnetic gating model is not met.  
Such a low value of the magnetic field strength, however is compatible with the centrifugal gating and quasi-spherical 
settling accretion models.\\

\textit{Acknowledgements:} This work was supported in part under NASA Contract No. NNG08FD60C, and made use of data from the \nustar\ mission, a project led by the California Institute of Technology, managed by the Jet Propulsion Laboratory, and funded by the National Aeronautics and Space Administration. We thank the \nustar\ Operations, Software and Calibration teams as well as the \sw\ team for support with the execution and analysis of these observations. This research has made use of the \nustar\ Data Analysis Software (\software{NuSTARDAS}) jointly developed by the ASI Science Data Center (ASDC, Italy) and the California Institute of Technology (USA). PR acknowledges contract ASI-INAF I/004/11/0. LN wishes to acknowledge the Italian Space Agency (ASI) for Financial support by ASI/INAF grant I/037/12/0-011/13. VB thanks Dipankar Bhattacharya for helpful discussions.

\textit{Facilities:} \nustar, \sw

\bibliographystyle{mn2e} 
\bibliography{igrj17544-2619}

\begin{thebibliography}{}
\makeatletter
\relax
\def\mn@urlcharsother{\let\do\@makeother \do\$\do\&\do\#\do\^\do\_\do\%\do\~}
\def\mn@doi{\begingroup\mn@urlcharsother \@ifnextchar [ {\mn@doi@}
  {\mn@doi@[]}}
\def\mn@doi@[#1]#2{\def\@tempa{#1}\ifx\@tempa\@empty \href
  {http://dx.doi.org/#2} {doi:#2}\else \href {http://dx.doi.org/#2} {#1}\fi
  \endgroup}
\def\mn@eprint#1#2{\mn@eprint@#1:#2::\@nil}
\def\mn@eprint@arXiv#1{\href {http://arxiv.org/abs/#1} {{\tt arXiv:#1}}}
\def\mn@eprint@dblp#1{\href {http://dblp.uni-trier.de/rec/bibtex/#1.xml}
  {dblp:#1}}
\def\mn@eprint@#1:#2:#3:#4\@nil{\def\@tempa {#1}\def\@tempb {#2}\def\@tempc
  {#3}\ifx \@tempc \@empty \let \@tempc \@tempb \let \@tempb \@tempa \fi \ifx
  \@tempb \@empty \def\@tempb {arXiv}\fi \@ifundefined
  {mn@eprint@\@tempb}{\@tempb:\@tempc}{\expandafter \expandafter \csname
  mn@eprint@\@tempb\endcsname \expandafter{\@tempc}}}

\bibitem[\protect\citeauthoryear{Bhalerao}{Bhalerao}{2012}]{bhalerao12}
Bhalerao V.~B.,  2012, PhD thesis, California Institute of Technology, \url
  {http://resolver.caltech.edu/CaltechTHESIS:05312012-150403422}

\bibitem[\protect\citeauthoryear{{Bozzo}, {Falanga}  \& {Stella}}{{Bozzo}
  et~al.}{2008}]{Bozzo2008}
{Bozzo} E.,  {Falanga} M.,   {Stella} L.,  2008, \mn@doi [\apj]
  {10.1086/589990}, \href {http://adsabs.harvard.edu/abs/2008ApJ...683.1031B}
  {683, 1031}

\bibitem[\protect\citeauthoryear{Caballero \& Wilms}{Caballero \&
  Wilms}{2012}]{cw12}
Caballero I.,  Wilms J.,  2012, Memorie della Societa Astronomica Italiana

\bibitem[\protect\citeauthoryear{{Clark}, {Hill}, {Bird}, {McBride}, {Scaringi}
   \& {Dean}}{{Clark} et~al.}{2009}]{Clark2009:17544-2619period}
{Clark} D.~J.,  {Hill} A.~B.,  {Bird} A.~J.,  {McBride} V.~A.,  {Scaringi} S.,
   {Dean} A.~J.,  2009, \mn@doi [\mnras] {10.1111/j.1745-3933.2009.00737.x},
  \href {http://adsabs.harvard.edu/abs/2009MNRAS.399L.113C} {399, L113}

\bibitem[\protect\citeauthoryear{{Coburn}, {Heindl}, {Rothschild}, {Gruber},
  {Kreykenbohm}, {Wilms}, {Kretschmar}  \& {Staubert}}{{Coburn}
  et~al.}{2002}]{Coburn2002}
{Coburn} W.,  {Heindl} W.~A.,  {Rothschild} R.~E.,  {Gruber} D.~E.,
  {Kreykenbohm} I.,  {Wilms} J.,  {Kretschmar} P.,   {Staubert} R.,  2002,
  \mn@doi [\apj] {10.1086/343033}, \href
  {http://adsabs.harvard.edu/abs/2002ApJ...580..394C} {580, 394}

\bibitem[\protect\citeauthoryear{{Drave}, {Bird}, {Townsend}, {Hill},
  {McBride}, {Sguera}, {Bazzano}  \& {Clark}}{{Drave}
  et~al.}{2012}]{Drave2012:17544_2619_pulsation}
{Drave} S.~P.,  {Bird} A.~J.,  {Townsend} L.~J.,  {Hill} A.~B.,  {McBride}
  V.~A.,  {Sguera} V.,  {Bazzano} A.,   {Clark} D.~J.,  2012, \mn@doi [\aap]
  {10.1051/0004-6361/201117947}, \href
  {http://adsabs.harvard.edu/abs/2012A%26A...539A..21D} {539, A21}

\bibitem[\protect\citeauthoryear{{Drave}, {Bird}, {Sidoli}, {Sguera},
  {Bazzano}, {Hill}  \& {Goossens}}{{Drave} et~al.}{2014}]{Drave2014:17544}
{Drave} S.~P.,  {Bird} A.~J.,  {Sidoli} L.,  {Sguera} V.,  {Bazzano} A.,
  {Hill} A.~B.,   {Goossens} M.~E.,  2014, \mn@doi [\mnras]
  {10.1093/mnras/stu110}, \href
  {http://adsabs.harvard.edu/abs/2014MNRAS.439.2175D} {439, 2175}

\bibitem[\protect\citeauthoryear{D’A\`{\i}, Cusumano, {La Parola}, Segreto,
  {Di Salvo}, Iaria  \& Robba}{D’A\`{\i} et~al.}{2011}]{dcl+11}
D’A\`{\i} A.,  Cusumano G.,  {La Parola} V.,  Segreto A.,  {Di Salvo} T.,
  Iaria R.,   Robba N.~R.,  2011, \mn@doi [Astronomy \& Astrophysics]
  {10.1051/0004-6361/201117035}, 532, A73

\bibitem[\protect\citeauthoryear{{Farinelli} et~al.,}{{Farinelli}
  et~al.}{2012}]{Farinelli2012:sfxts_paperVIII}
{Farinelli} R.,  et~al., 2012, \mn@doi [\mnras]
  {10.1111/j.1365-2966.2012.21422.x}, \href
  {http://adsabs.harvard.edu/abs/2012MNRAS.424.2854F} {424, 2854}

\bibitem[\protect\citeauthoryear{Frank, King  \& Raine}{Frank
  et~al.}{2002}]{fkr02}
Frank J.,  King A.,   Raine D.~J.,  2002, Accretion Power in Astrophysics

\bibitem[\protect\citeauthoryear{Grebenev}{Grebenev}{2009}]{grebenev09}
Grebenev S.,  2009, "Proceedings of The Extreme sky: Sampling the Universe
  above 10 keV. October 13-17 2009

\bibitem[\protect\citeauthoryear{{Grebenev} \& {Sunyaev}}{{Grebenev} \&
  {Sunyaev}}{2007}]{Grebenev2007}
{Grebenev} S.~A.,  {Sunyaev} R.~A.,  2007, \mn@doi [Astronomy Letters]
  {10.1134/S1063773707030024}, \href
  {http://adsabs.harvard.edu/abs/2007AstL...33..149G} {33, 149}

\bibitem[\protect\citeauthoryear{{Grebenev}, {Lutovinov}  \&
  {Sunyaev}}{{Grebenev} et~al.}{2003}]{Grebenev2003:17544-2619}
{Grebenev} S.~A.,  {Lutovinov} A.~A.,   {Sunyaev} R.~A.,  2003, \ATel, \href
  {http://adsabs.harvard.edu/abs/2003ATel..192....1G} {192, 1}

\bibitem[\protect\citeauthoryear{{Grebenev}, {Rodriguez}, {Westergaard},
  {Sunyaev}  \& {Oosterbroek}}{{Grebenev}
  et~al.}{2004}]{Grebenev2004:17544-2619}
{Grebenev} S.~A.,  {Rodriguez} J.,  {Westergaard} N.~J.,  {Sunyaev} R.~A.,
  {Oosterbroek} T.,  2004, \ATel, \href
  {http://adsabs.harvard.edu/abs/2004ATel..252....1G} {252, 1}

\bibitem[\protect\citeauthoryear{Ho \& Lai}{Ho \& Lai}{2003}]{hl03}
Ho W. C.~G.,  Lai D.,  2003, \mn@doi [\mnras]
  {10.1046/j.1365-8711.2003.06047.x}, 338, 233

\bibitem[\protect\citeauthoryear{{Krimm} et~al.,}{{Krimm}
  et~al.}{2007}]{Krimm2007:ATel1265}
{Krimm} H.~A.,  et~al., 2007, \ATel, \href
  {http://adsabs.harvard.edu/abs/2007ATel.1265....1K} {1265, 1}

\bibitem[\protect\citeauthoryear{{Kuulkers} et~al.,}{{Kuulkers}
  et~al.}{2007}]{Kuulkers2007}
{Kuulkers} E.,  et~al., 2007, \ATel, \href
  {http://adsabs.harvard.edu/abs/2007ATel.1266....1K} {1266, 1}

\bibitem[\protect\citeauthoryear{{Liu}, {van Paradijs}  \& {van den
  Heuvel}}{{Liu} et~al.}{2005}]{Liu2005:hmxb_LMC_SMC}
{Liu} Q.~Z.,  {van Paradijs} J.,   {van den Heuvel} E.~P.~J.,  2005, \mn@doi
  [\aap] {10.1051/0004-6361:20053718}, \href
  {http://adsabs.harvard.edu/abs/2005A%26A...442.1135L} {442, 1135}

\bibitem[\protect\citeauthoryear{{Liu}, {van Paradijs}  \& {van den
  Heuvel}}{{Liu} et~al.}{2006}]{Liu2006:hmxb_Gal}
{Liu} Q.~Z.,  {van Paradijs} J.,   {van den Heuvel} E.~P.~J.,  2006, \mn@doi
  [\aap] {10.1051/0004-6361:20064987}, \href
  {http://adsabs.harvard.edu/abs/2006A%26A...455.1165L} {455, 1165}

\bibitem[\protect\citeauthoryear{{Negueruela}, {Smith}, {Reig}, {Chaty}  \&
  {Torrej{\'o}n}}{{Negueruela} et~al.}{2006a}]{Negueruela2006:ESASP604}
{Negueruela} I.,  {Smith} D.~M.,  {Reig} P.,  {Chaty} S.,   {Torrej{\'o}n}
  J.~M.,  2006a, in {A.~Wilson} ed.,  ESA Special Publication Vol. 604, The
  X-ray Universe 2005. p.~165, \mn@eprint {} {arXiv:astro-ph/0511088}

\bibitem[\protect\citeauthoryear{{Negueruela}, {Smith}, {Harrison}  \&
  {Torrej{\'o}n}}{{Negueruela} et~al.}{2006b}]{Negueruela2006}
{Negueruela} I.,  {Smith} D.~M.,  {Harrison} T.~E.,   {Torrej{\'o}n} J.~M.,
  2006b, \mn@doi [\apj] {10.1086/498935}, \href
  {http://adsabs.harvard.edu/abs/2006ApJ...638..982N} {638, 982}

\bibitem[\protect\citeauthoryear{{Negueruela}, {Torrej{\'o}n}, {Reig},
  {Rib{\'o}}  \& {Smith}}{{Negueruela} et~al.}{2008}]{Negueruela2008}
{Negueruela} I.,  {Torrej{\'o}n} J.~M.,  {Reig} P.,  {Rib{\'o}} M.,   {Smith}
  D.~M.,  2008, in {Bandyopadhyay} R.~M.,  {Wachter} S.,  {Gelino} D.,
  {Gelino} C.~R.,  eds,  American Institute of Physics Conference Series Vol.
  1010, A Population Explosion: The Nature \& Evolution of X-ray Binaries in
  Diverse Environments. pp 252--256, \mn@eprint {arXiv} {0801.3863},
  \mn@doi{10.1063/1.2945052}

\bibitem[\protect\citeauthoryear{{Pellizza}, {Chaty}  \&
  {Negueruela}}{{Pellizza} et~al.}{2006}]{Pellizza2006}
{Pellizza} L.~J.,  {Chaty} S.,   {Negueruela} I.,  2006, \mn@doi [\aap]
  {10.1051/0004-6361:20054436}, \href
  {http://adsabs.harvard.edu/abs/2006A%26A...455..653P} {455, 653}

\bibitem[\protect\citeauthoryear{{Rahoui}, {Chaty}, {Lagage}  \&
  {Pantin}}{{Rahoui} et~al.}{2008a}]{Rahoui2008}
{Rahoui} F.,  {Chaty} S.,  {Lagage} P.-O.,   {Pantin} E.,  2008a, \mn@doi
  [\aap] {10.1051/0004-6361:20078774}, \href
  {http://adsabs.harvard.edu/abs/2008A%26A...484..801R} {484, 801}

\bibitem[\protect\citeauthoryear{Rahoui, Chaty, Lagage  \& Pantin}{Rahoui
  et~al.}{2008b}]{rcl+08}
Rahoui F.,  Chaty S.,  Lagage P.-O.,   Pantin E.,  2008b, \mn@doi [Astronomy
  and Astrophysics] {10.1051/0004-6361:20078774}, 484, 801

\bibitem[\protect\citeauthoryear{{Rampy}, {Smith}  \& {Negueruela}}{{Rampy}
  et~al.}{2009}]{Rampy2009:suzaku17544}
{Rampy} R.~A.,  {Smith} D.~M.,   {Negueruela} I.,  2009, \mn@doi [\apj]
  {10.1088/0004-637X/707/1/243}, \href
  {http://adsabs.harvard.edu/abs/2009ApJ...707..243R} {707, 243}

\bibitem[\protect\citeauthoryear{{Reig}}{{Reig}}{2011}]{Reig2011}
{Reig} P.,  2011, \mn@doi [\apss] {10.1007/s10509-010-0575-8}, \href
  {http://adsabs.harvard.edu/abs/2011Ap%26SS.332....1R} {332, 1}

\bibitem[\protect\citeauthoryear{{Romano}, {Sidoli}, {Mangano}, {Mereghetti}
  \& {Cusumano}}{{Romano} et~al.}{2007}]{Romano2007}
{Romano} P.,  {Sidoli} L.,  {Mangano} V.,  {Mereghetti} S.,   {Cusumano} G.,
  2007, \mn@doi [\aap] {10.1051/0004-6361:20077383}, \href
  {http://adsabs.harvard.edu/abs/2007A%26A...469L...5R} {469, L5}

\bibitem[\protect\citeauthoryear{Romano et~al.,}{Romano et~al.}{2011a}]{rlv+11}
Romano P.,  et~al., 2011a, \mn@doi [\mnras] {10.1111/j.1365-2966.2010.17564.x},
  410, 1825

\bibitem[\protect\citeauthoryear{{Romano} et~al.,}{{Romano}
  et~al.}{2011b}]{Romano2011:sfxts_paperVII}
{Romano} P.,  et~al., 2011b, \mn@doi [\mnras]
  {10.1111/j.1745-3933.2010.00999.x}, \href
  {http://adsabs.harvard.edu/abs/2011MNRAS.412L..30R} {412, L30}

\bibitem[\protect\citeauthoryear{{Romano} et~al.,}{{Romano}
  et~al.}{2013}]{Romano2013:Cospar12}
{Romano} P.,  et~al., 2013, \mn@doi [Advances in Space Research]
  {10.1016/j.asr.2013.07.034}, \href
  {http://cdsads.u-strasbg.fr/abs/2013AdSpR..52.1593R} {52, 1593}

\bibitem[\protect\citeauthoryear{{Romano} et~al.,}{{Romano}
  et~al.}{2014}]{Romano2014:sfxts_catI}
{Romano} P.,  et~al., 2014, \mn@doi [\aap] {10.1051/0004-6361/201322516}, \href
  {http://adsabs.harvard.edu/abs/2014A%26A...562A...2R} {562, A2}

\bibitem[\protect\citeauthoryear{{Sguera} et~al.,}{{Sguera}
  et~al.}{2005}]{Sguera2005}
{Sguera} V.,  et~al., 2005, \mn@doi [\aap] {10.1051/0004-6361:20053103}, \href
  {http://adsabs.harvard.edu/cgi-bin/nph-bib_query?bibcode=2005A%26A...444..221S&db_key=AST}
  {444, 221}

\bibitem[\protect\citeauthoryear{{Sguera} et~al.,}{{Sguera}
  et~al.}{2006}]{Sguera2006}
{Sguera} V.,  et~al., 2006, \mn@doi [\apj] {10.1086/504827}, \href
  {http://adsabs.harvard.edu/cgi-bin/nph-bib_query?bibcode=2006ApJ...646..452S&db_key=AST}
  {646, 452}

\bibitem[\protect\citeauthoryear{{Shakura}, {Postnov}, {Sidoli}  \&
  {Paizis}}{{Shakura} et~al.}{2014}]{Shakura2014:bright_flares}
{Shakura} N.,  {Postnov} K.,  {Sidoli} L.,   {Paizis} A.,  2014, \mn@doi
  [\mnras] {10.1093/mnras/stu1027}, \href
  {http://adsabs.harvard.edu/abs/2014MNRAS.442.2325S} {442, 2325}

\bibitem[\protect\citeauthoryear{{Sidoli}, {Romano}, {Mereghetti}, {Paizis},
  {Vercellone}, {Mangano}  \& {G{\"o}tz}}{{Sidoli} et~al.}{2007}]{Sidoli2007}
{Sidoli} L.,  {Romano} P.,  {Mereghetti} S.,  {Paizis} A.,  {Vercellone} S.,
  {Mangano} V.,   {G{\"o}tz} D.,  2007, \mn@doi [\aap]
  {10.1051/0004-6361:20078137}, \href
  {http://adsabs.harvard.edu/abs/2007A%26A...476.1307S} {476, 1307}

\bibitem[\protect\citeauthoryear{{Sidoli} et~al.,}{{Sidoli}
  et~al.}{2009a}]{Sidoli2009:sfxts_paperIV}
{Sidoli} L.,  et~al., 2009a, \mn@doi [\mnras]
  {10.1111/j.1365-2966.2009.15049.x}, 397, 1528

\bibitem[\protect\citeauthoryear{{Sidoli} et~al.,}{{Sidoli}
  et~al.}{2009b}]{Sidoli2009:sfxts_paperIII}
{Sidoli} L.,  et~al., 2009b, \mn@doi [\apj] {10.1088/0004-637X/690/1/120},
  \href {http://cdsads.u-strasbg.fr/abs/2009ApJ...690..120S} {690, 120}

\bibitem[\protect\citeauthoryear{Smith}{Smith}{2014}]{smith14}
Smith D.~M.,  2014, The Astronomer's Telegram

\bibitem[\protect\citeauthoryear{{Smith}, {Negueruela}, {Heindl}, {Markwardt}
  \& {Swank}}{{Smith} et~al.}{2004}]{Smith2004:fast_transients}
{Smith} D.~M.,  {Negueruela} I.,  {Heindl} W.~A.,  {Markwardt} C.~B.,   {Swank}
  J.~H.,  2004, in \baas. p.~954

\bibitem[\protect\citeauthoryear{{Sunyaev}, {Grebenev}, {Lutovinov},
  {Rodriguez}, {Mereghetti}, {Gotz}  \& {Courvoisier}}{{Sunyaev}
  et~al.}{2003}]{Sunyaev2003}
{Sunyaev} R.~A.,  {Grebenev} S.~A.,  {Lutovinov} A.~A.,  {Rodriguez} J.,
  {Mereghetti} S.,  {Gotz} D.,   {Courvoisier} T.,  2003, \ATel, \href
  {http://adsabs.harvard.edu/abs/2003ATel..190....1S} {190, 1}

\bibitem[\protect\citeauthoryear{Tiengo et~al.,}{Tiengo et~al.}{2013}]{tem+13}
Tiengo A.,  et~al., 2013, \mn@doi [Nature] {10.1038/nature12386}, 500, 312

\bibitem[\protect\citeauthoryear{Verner, Ferland, Korista  \& Yakovlev}{Verner
  et~al.}{1996}]{vfk+96}
Verner D.~A.,  Ferland G.~J.,  Korista K.~T.,   Yakovlev D.~G.,  1996, \mn@doi
  [The Astrophysical Journal] {10.1086/177435}, 465, 487

\bibitem[\protect\citeauthoryear{{Walter} \& {Zurita Heras}}{{Walter} \&
  {Zurita Heras}}{2007}]{Walter2007}
{Walter} R.,  {Zurita Heras} J.,  2007, \mn@doi [\aap]
  {10.1051/0004-6361:20078353}, \href
  {http://adsabs.harvard.edu/abs/2007A%26A...476..335W} {476, 335}

\bibitem[\protect\citeauthoryear{Wilms, Allen  \& McCray}{Wilms
  et~al.}{2000}]{wam00}
Wilms J.,  Allen A.,   McCray R.,  2000, \mn@doi [The Astrophysical Journal]
  {10.1086/317016}, 542, 914

\bibitem[\protect\citeauthoryear{{in't Zand}}{{in't Zand}}{2005}]{zand2005}
{in't Zand} J.~J.~M.,  2005, \mn@doi [\aap] {10.1051/0004-6361:200500162},
  \href {http://adsabs.harvard.edu/abs/2005A%26A...441L...1I} {441, L1}

\bibitem[\protect\citeauthoryear{{in't Zand}, {Heise}, {Ubertini}, {Bazzano}
  \& {Markwardt}}{{in't Zand} et~al.}{2004}]{zand2004:17544bepposax}
{in't Zand} J.,  {Heise} J.,  {Ubertini} P.,  {Bazzano} A.,   {Markwardt} C.,
  2004, in {V.~Schoenfelder, G.~Lichti, \& C.~Winkler} ed.,  ESA Special
  Publication Vol. 552, 5th INTEGRAL Workshop on the INTEGRAL Universe. p.~427

\makeatother
\end{thebibliography}

\label{lastpage}

\end{document}